\documentclass[12pt]{article}
\usepackage{graphicx} 
\begin{document}

\centerline{\bf Tax evasion dynamics and Zaklan model on Opinion-dependent Network$^{\star}$ }

\bigskip
\centerline{ F. W. S. Lima}
\bigskip

\noindent
Dietrich Stauffer Computational Physics Lab, Departamento de F\'{\i}sica, \\
Universidade Federal do Piau\'{\i}, 64049-550, Teresina - PI, Brazil.\\
\medskip

e-mail: fwslima@gmail.com

\bigskip

\hspace{4.7pc}{$\star$ This paper is dedicated to Dietrich Stauffer}\\

\begin{abstract}
Within the context of agent-based Monte-Carlo simulations, we study the well-known majority-vote model (MVM) with noise applied to tax evasion on Stauffer-Hohnisch-Pittnauer (SHP)  networks. To control the fluctuations for tax evasion in the economics model proposed by Zaklan, MVM is applied  in the  neighborhood of the critical noise $q_{c}$ to evolve the Zaklan model. The Zaklan model  had been studied recently using the equilibrium Ising model. Here we show that the Zaklan model is robust because this can be studied besides using equilibrium dynamics of Ising model also through the nonequilibrium MVM and on various topologies giving the same behavior regardless of dynamic or topology used here.  \\
Keywords: Opinion dynamics, Sociophysics, Majority vote, Nonequilibrium.
\end{abstract}

\section{Introduction}
The Ising model \cite{a3,onsager} has become a excellent tool to study other models of social application. The Ising model was already applied decades ago to explain how a school of fish aligns into one direction for swimming \cite{callen} or how workers decide whether or not to go on strike \cite{galam}. In the Latan\'e model of Social Impact \cite{latané} the Ising model has been used to give a consensus, a fragmentation into many different opinions, or a leadership effect when a few people change the opinion of lots of others. To some extent the voter model of Liggett \cite{ligg} is  a zero-temperature Ising-type model: opinions follow the majority of the neighbourhood, similar to Schelling \cite{Schel}, all these cited models and others can be found out in \cite{book}. Already F\"ollmer (1974) \cite{foo} applied the Ising
model to economics. 

The tax evasion remains to be a major predicament facing governments \cite{K,JA,L,JS}. Experimental evidence provided by G\"achter \cite{Ga} indeed suggests that tax payers tend to condition their decision regarding whether to pay taxes or not on the tax evasion decision of the members of their group. Frey and Torgler \cite{FT} also provide empirical evidence on the relevance of conditional cooperation for tax morale. Following the same context, recently, Zaklan et al. \cite{zaklan,zaklan1} developed an economics model to study the problem of tax evasion dynamics using the Ising model through Monte-Carlo simulations with the Glauber and heatbath algorithms (that obey detailed-balance equilibrium) to study the proposed model.

Grinstein et al. \cite{g} have argued that nonequilibrium stochastic spin systems on regular square lattices with up-down symmetry fall into the universality
class of the equilibrium Ising model \cite{g}. This conjecture was confirmed for various Archimedean lattices and 
in several models that do not obey detailed balance \cite{C,J,M,mario,lima01}. The majority-vote model (MVM) is a nonequilibrium model proposed by M.J. Oliveira in $1992$ and defined by stochastic dynamics with local rules and with up-down symmetry on a regular lattice shows a second-order phase transition with critical exponents $\beta$, $\gamma$, $\nu$ which characterize  the system in the vicinity of the phase transition identical \cite{mario,a1} with those of the equilibrim Ising model \cite{a3} for regular lattices. Lima et al. \cite{lima0} 
studied MVM  on VD random lattices
with periodic boundary conditions. These lattices posses natural quenched
disorder in their connections. They showed that the presence of quenched 
connectivity disorder is enough to alter the exponents $\beta/\nu$
and $\gamma/\nu$ from the pure model and therefore is a relevant term to
such non-equilibrium phase-transition in disagreement with the arguments of Grinstein et al. \cite{g}. Recently, simulations on both {\it undirected} and {\it directed} scale-free 
networks \cite{newman,sanchez,ba1,alex,sumour,sumourss,lima}, random graphs \cite{erdo,er2} and social networks \cite{er,er1,DS}, have attracted interest of researchers from various areas. These complex networks have been 
studied extensively by Lima et al. in the context of magnetism (MVM, Ising, and Potts model) \cite{lima1,lima2,lima3,lima4,lima5,lima6}, econophysics models \cite{zaklan1,lima8,limanew} and sociophysics model \cite{lima9, WGa}. 

In the present work, we study the behavior of the tax evasion on SHP networks using the dynamics of MVM, furthermore add a policy makers's tax enforcement mechanism consisting of two components: a probability of an audit each person is subject to in every period and a length of time detected tax evaders remain honest. We aim here is to extend the study of Zaklan et al. \cite{zaklan,zaklan1}, which illustrates how different levels of enforcement affect the tax evasion over time, to dynamics of MVM as an alternative model of  nonequilibrium to the Ising model that is capable of reproduce the same results for analysis and control of the tax evasion fluctuations. Then, we show that the Zaklan model is very robust for equilibrium and nonequilibrium models and also for various topologies used here. We show that the choice of using the Ising (equilibrium dynamics) or MVM (nonequilibrium dynamics) used to evolve
the Zaklan model is irrelevant, because the results obtained in this work are about the same for both Ising and MVM. The Zaklan model also is robust, because it works on SHP network. We show that for this topology the Zaklan model reaches our objective, that is, to  
control the tax evasion of a country (Germany and others). This does not occur with other models as Axelrod-Ross model for evolution of ethnocentrism \cite{lima9}, because the results are different depending of the topology of the network. The Ising model also is not robust, because on directed Barab\'asi-Albert network there is no  phase transition present as also on directed SL, $3D$, $4D$ and directed hypercubic lattices \cite{lima}. As described above, the MVM was
proposed by M.J. Oliveira in $1992$ \cite{mario} in order to improve the criterion of Grinstein et al. \cite{g}, initially described above. In the order to achieve his goal he used $4^{4}$ (SL) Archimedean lattice. However, also with the aim to improve this criterion other researchers studied MVM on several other topologies that are not Archimedeans \cite{campos, lima0,lima2, pereira, lima2, edina, adriano}. To their surprise all results obtained for the critical exponents are different from results obtained by M.J. Oliveira,
and are also different for each topology used. Pereira et al. \cite{pereira} then concluded that MVM has different universality classes which depend only on the topology used, and that all have one thing in common that is their
$2\beta/\nu+ \gamma/\nu=D_{eff}$ effective dimension, obtained by critical exponents for each topology used, equals $D_{eff}=1$. Here, we show that the Zaklan model behavior is similar for all topologies or dynamics studied here. Therefore, we believe that this model is very robust, different from the other models cited above. Galam \cite{galam1,galam2,galam3,galam4,galam5,galam6} introduced for the first time local majority rules in social systems to the field of sociophysics using discrete opinion models. Here, we also hope to introduce  for the first time the use of MVM to the field of sociophysics or econophysics using discrete opinions as in the Zaklan model. Therefore, we do not live in a social equilibrium, any rumor or gossip can lead to a government or market chaos and we believe that nothing is better than a non­equilibrium model (MVM) to explain events  of non­equilibrium. The difference of this work compared to previous studies using the Zaklan's model this is the application on the social networks that follow the prescription of Hohnisch bonds of SHP networks \cite{DS}. 

Hohnisch bonds of SHP \cite{DS} are links connecting nodes with different values
(spins, opinions, ...) on them; they are at each time step with a low 
probability 0.0001 replaced by a link to another randomly selected node. Links
connecting agreeing nodes are not replaced. In the present work we start with
each node having links to four randomly selected neighbors. All links are directed. 

The remainder of our paper is organised as follows. In section 2, we present the Zaklan model evolving with dynamics of MVM. In section 3 we make an analysis of tax evasion dynamics with the Zaklan model on two-dimensional square lattices using 
MVM for their temporal evolution under different enforcement regimes; we discuss the results obtained. In section 4 we show that MVM also is capable to control the different levels of the tax evasion analysed in section 3, as it was made by Zaklan et al. \cite{zaklan1} using Ising models. We use the enforcement mechanism cited above on  SHP network, we discuss the resulting tax evasion dynamics. Finally in section 5 we present our conclusions about the study of the Zaklan model using MVM. Results for different topologies (directed and undirected Barab\'asi-Albert networks, simple square lattices) were given in \cite{limanew}. 

\section{Zaklan model }

On SHP networks each site of the network is inhabited, at each time step, by an  agent with "voters" or spin variables ${\sigma}$ taking the values $+1$ representing an honest tax payer, or $-1$ trying to at least partially escape her tax duty. Here is assumed that initially everybody is honest. Each period individuals can rethink their behavior and have the opportunity to become the opposite type of agent they were in previous period. In each time period the system evolves by a single spin-flip dynamics with a probability $w_{i}$ given by
\begin{equation}
w_{i}(\sigma)=\frac{1}{2}\biggl[ 1-(1-2q)\sigma_{i}S\biggl(\sum_{\delta
=1}^{k_{i}}\sigma_{i,\delta}\biggl)\biggl],
\end{equation}
where $S(x)$ is the sign $\pm 1$ of $x$ if $x\neq0$, $S(x)=0$ if $x=0$, and the 
summation runs over all $k_i$ nearest-neighbour sites $\sigma_{i,\delta}$ of $\sigma_{i}$. In this model an agent assumes the value $\pm 1$ depending on the opinion of the majority of its neighbors. The control noise parameter $q$ plays the role of the temperature in equilibrium systems and measures the probability of aligning $\sigma_{i}$ antiparallel to the majority of its neighbors $\sigma_{i,\delta}$.
Then various degrees of homogeneity regarding either opinion are possible. An extremely homogenous group is entirely made either of honest people or of tax evaders, depending of the sign $S(x)$ of the majority of neighbhors. If $S(x)$ of the neighbors is zero the agent $\sigma_{i}$ will be honest or evader in the next time period with probability $1/2$. We further introduce a probability of an efficient audit ($p$). Therefore, if tax evasion is detected, the agent must remain honest for a number $k$ of time steps. Here, one time step is one sweep through the entire network.

\section{Controlling the tax evasion dynamics}

In order to calculate the rate of tax evaders, we use the equation below,
\begin{equation}
{\rm tax \; evasion} =\biggl[N-N_{\rm honest}\biggl]/N,
\end{equation}
where $N$ is the total number and $N_{\rm honest}$ the honest number of agents. 
The tax evasion is calculated at every time step $t$ of system evolution; one
time step is one sweep through the entire network.

Here, we first will present the  baseline case $k=0$, i.e., no use of enforcement, at $q=0.95q_{c}$ and with $N=400$ sites
for SHP network. All simulation are performed over $20,000$ time steps, as shown in Fig. 1. For very low noises the part of autonomous decisions almost completely disappears. The individuals then base their decision solely on what most of their neighbours do. A rising noise has the opposite effect. Individuals then decide more autonomously. 
For MVM it is known that for $q > q_{c}$, half of the people are honest and the other half cheat, while for $q<q_{c}$  either one opinion or the other opinion dominates. Because of
this behavior we set at fixed "Social Temperature" ($q$) to some values slightly below $q_{c}$, where the case that agents distribute in equal proportions onto the two alternatives is excluded. Then having set the noise parameter $q$ close to $q_{c}=0.116$ on the SHP network,  we vary the degrees of punishment ($k=1$, $10$ and $50$) and audit probability 
rate ($p=0.5\%$, $10\%$ and $90\%$). Therefore, if tax evasion is detected, the  enforcement mechanism ($p$) and the period time of punishment $k$ are triggered in order to control the tax evasion level. The punished individuals 
remain honest for a certain number $k$ of periods, as explained before in section 2.
\newpage
\begin{figure}[ht]
\begin{center}
\includegraphics[angle=0,scale=0.5]{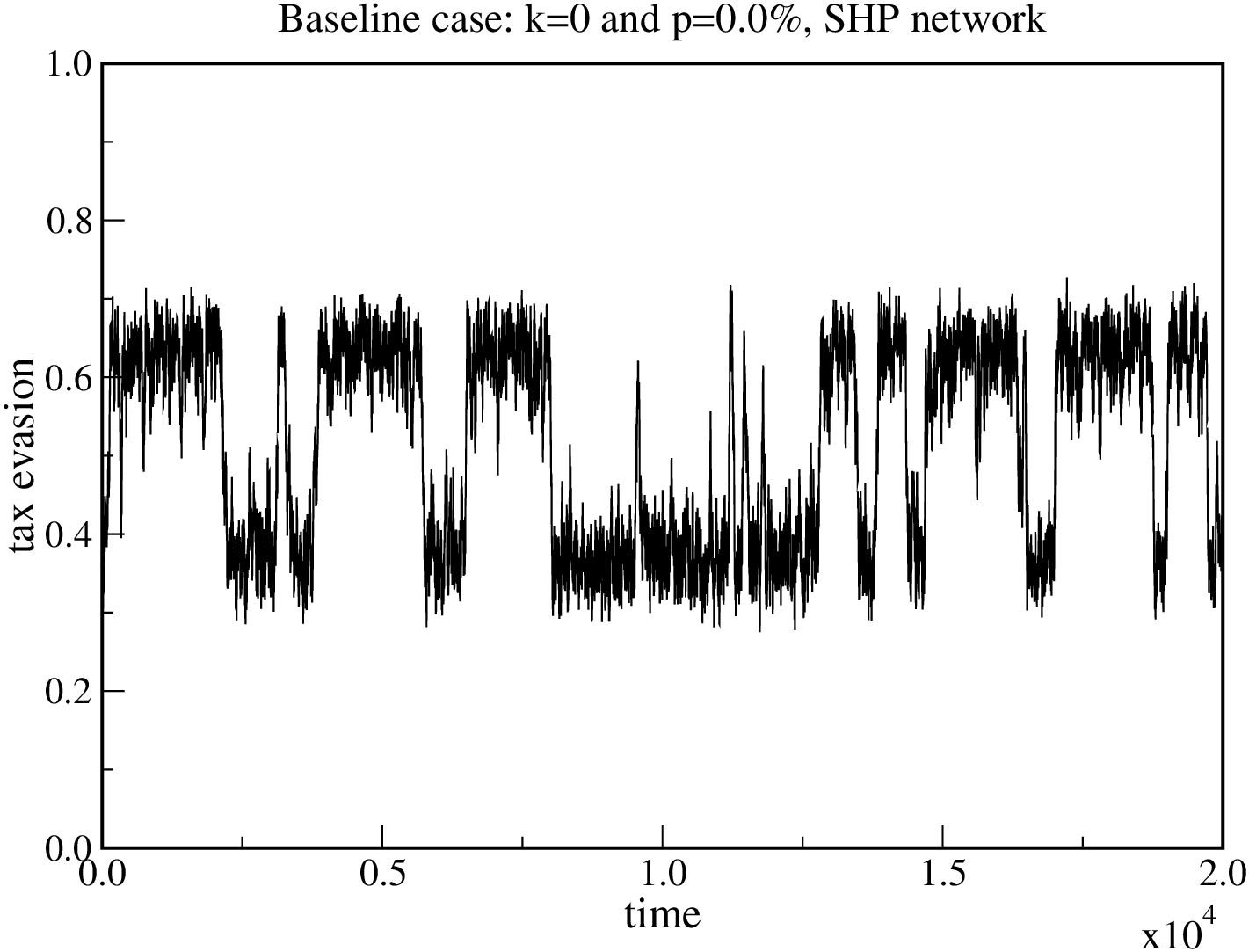}
\end{center}
\caption{Baseline case for SHP network. We use $q=0.95q_{c}$ on SHP networks and perform all simulation over $20,000$ time steps, also in the later figures.}
\end{figure} 

In Fig. 1 we plot the baseline case $k=0$, i.e., no use of enforcement, for the SHP for dynamics of the tax evasion over $20,000$ time steps. Although everybody is honest initially, it is impossible to predict roughly which level of tax compliance will be reached at some time step in the future.
\newpage
\begin{figure}[ht]
\begin{center}
\includegraphics[angle=0,scale=0.5]{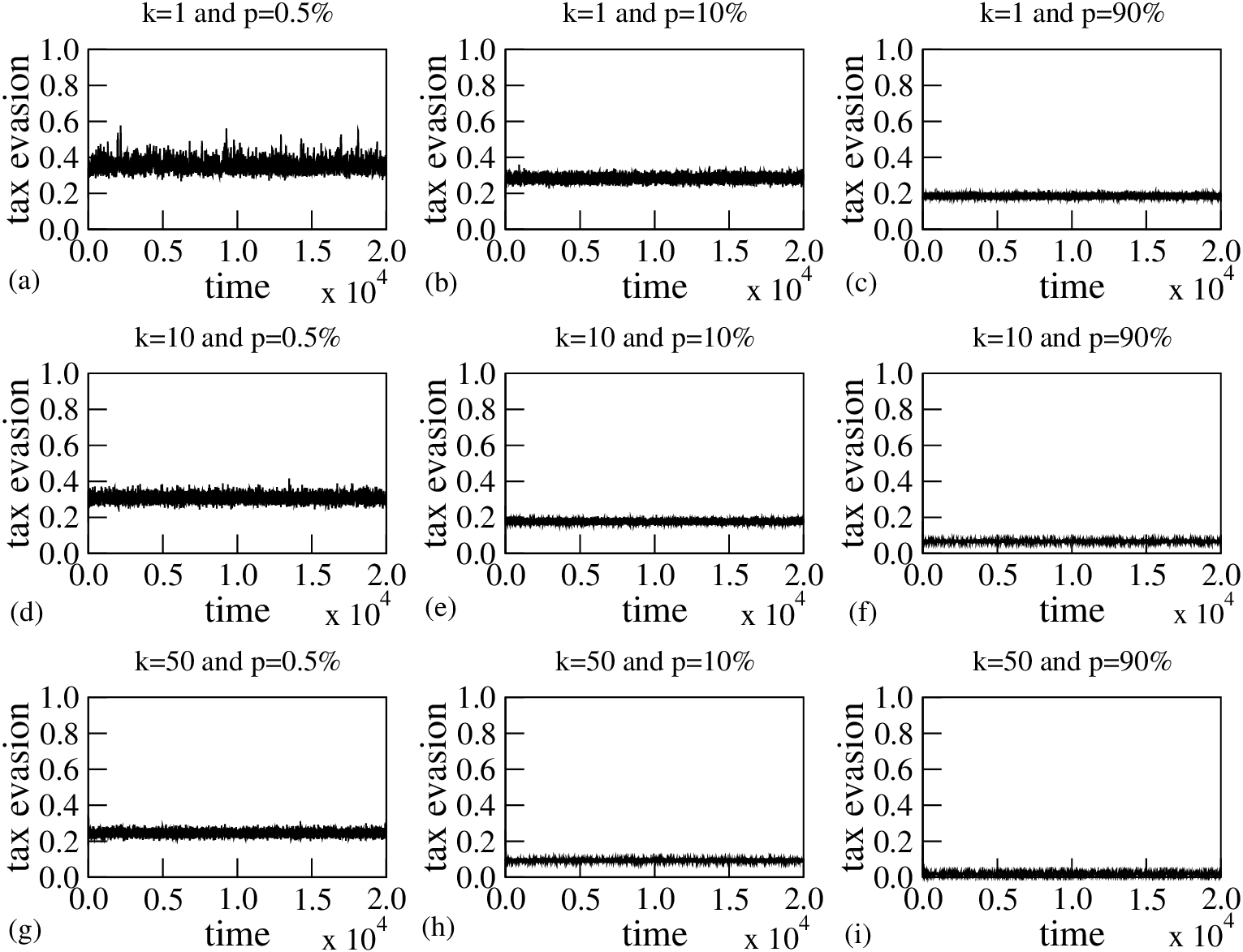}
\end{center}
\caption{The SHP networks of tax evasion with various degrees of enforcement.} 
\end{figure}
Figure 2 illustrates different simulation settings for SHP networks, for
each considered combination of degree of punishment ($k=1$, $10$ and $50$)
and audit probability ($p=0.5\%$, $10\%$ and $90\%$), where the tax evasion
is plotted over 20,000 time steps. Here we show that even a very small level
the enforcement ($p=0.5\%$ and $k=1$) suffices to reduce
fluctuations in tax evasion and to establish mainly compliance. Both a rise in
audit probability (greater $p$) and a higher penalty (greater $k$) work to
flatten the time series of tax evasion and to shift the band of possible
non-compliance values towards more compliance. However, the simulations
show that even extreme enforcement measures ($p=90\%$  and $k=50$)
cannot fully solve the problem of tax evasion.
\newpage
\begin{figure}[ht]
\begin{center}
\includegraphics[angle=0,scale=0.5]{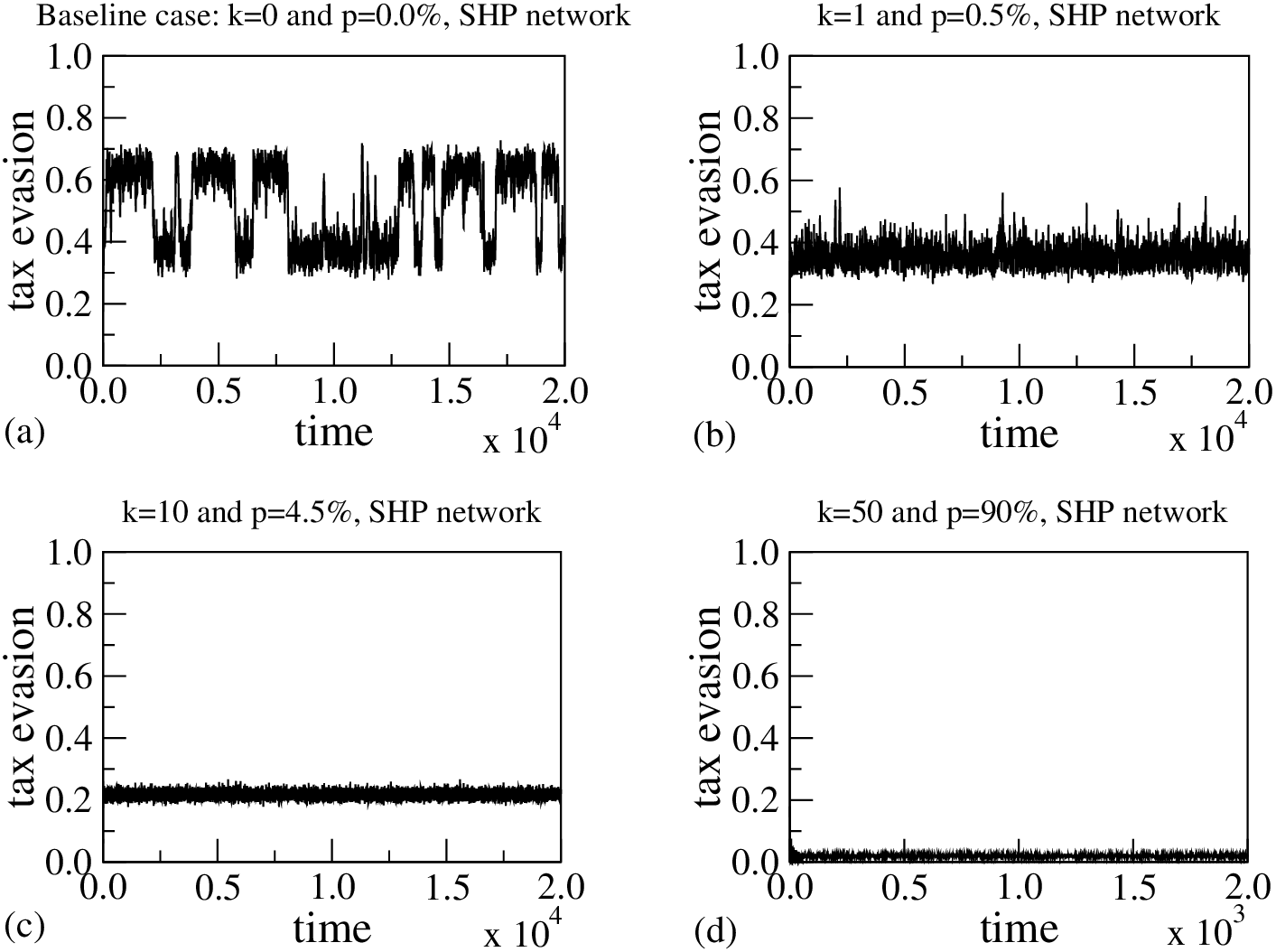}
\end{center}
\caption{Tax evasion for different enforcement regimes SHP Network and for degrees of punishment $k=0$, $1$, $10$ and $50$ and audit 
probability $p=0.0\%, 0.5\%, 4.5\%$, and $90\%$. }
\end{figure} 

In Fig. 3 we plot tax evasion for SHP network with $N=400$, again for different enforcement $k=0$, $k=1$, $10$, and $50$, but now with audit probability $p=0.0\%$, $p=0.5\%$, $p=4.5\%$, and $p=90\%$. For case (a) we plot the baseline case $k=0$, i.e., no use of enforcement for SHP networks and parameters as in Fig. 1. As everybody is honest initially, it is impossible to predict which level of tax compliance will be reached at some time step in the future. For case (b) we show the tax evasion level decreases with audit probability increasing to $p=0.5\%$  even for small of punishment 
$k=1$ . In the case (c) we show the tax evasion level decreases, on SHP network, for a more realistic set of possible values degrees of punishment $k=10$ and audit probability $p=4.5\%$ \cite{Ga, WGa, zaklan}. In the case (d) we also show tax evasion level decreases much more for an extreme set of punishment $k=50$ and audit probability $p=90\%$ \cite{Ga, zaklan}.
\newpage
\begin{figure}[ht]
\begin{center}
\includegraphics[angle=0,scale=0.5]{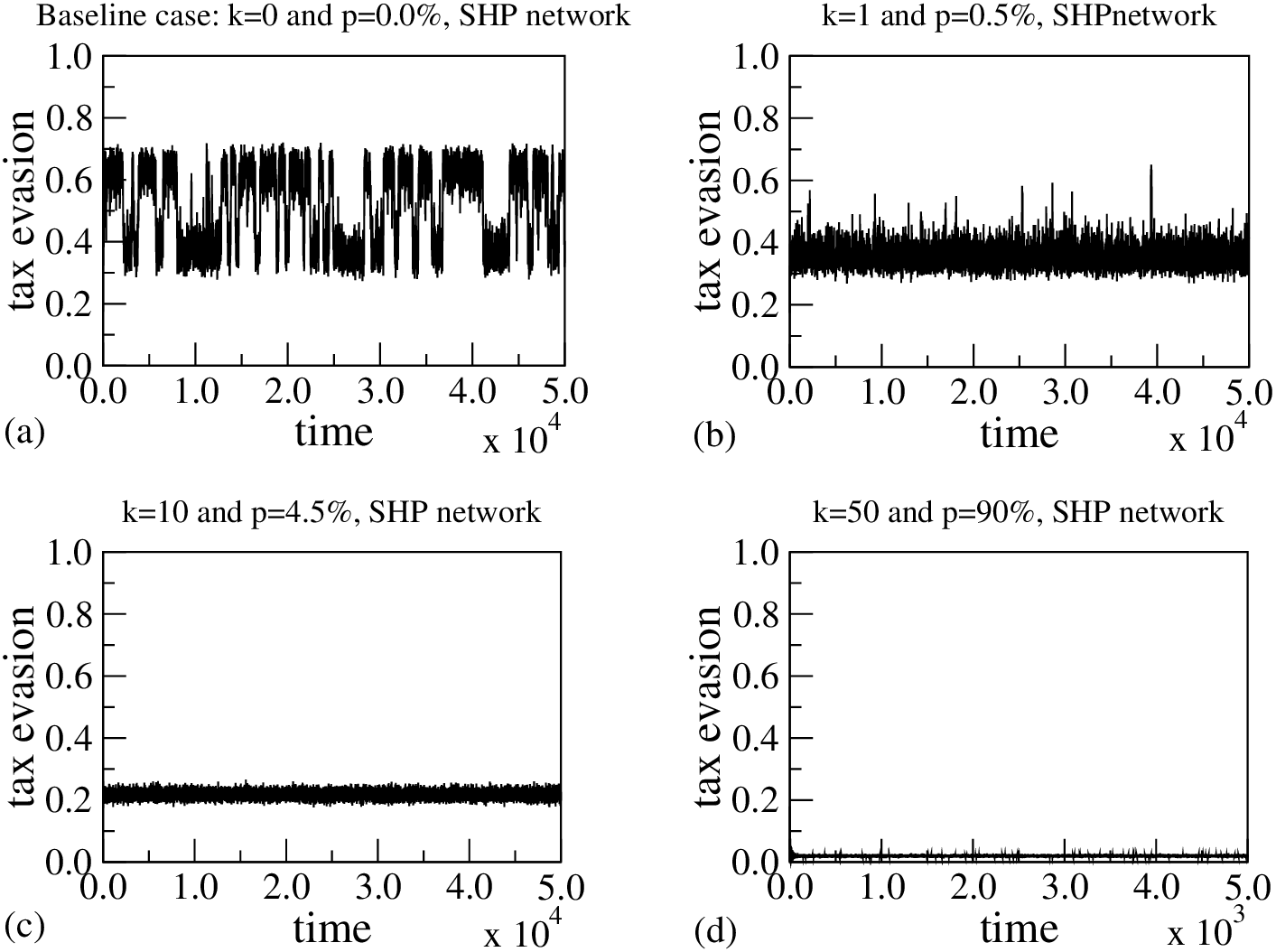}
\end{center}
\caption{ Tax evasion for different enforcement regimes SHP Network and for degrees of punishment $k=0$, $1$, $10$ and $50$ and audit probability 
$p=0.0\%, 0.5\%, 4.5\%$, and $90\%$. Now differently from our earlier figures we use $N=4,000$ 
sites (nodes) of SHP network.}
\end{figure}

In Fig. 4 we plot tax evasion for SHP network with $N=4,000$ (and not $N=400$ as for the figures before), again for the same parameters studied in the Fig. 3. Therefore, in Fig. 4 we showed that 
also for large networks with  $4,000$ nodes and $50,000$ time steps, studied here, the control of tax evasion is achieved by increasing $k$ and/or $p$.

\section{Conclusion}
Wintrobe and  G\"erxhani \cite{WGa} explain the observed higher level of tax evasion in generally less developed countries with a lower amount of trust that people have in governmental institutions. To study this problem Zaklan et al. \cite{zaklan,zaklan1} proposed a model, called here the Zaklan model, using Monte Carlo simulations and a equilibrium dynamics (Ising model) on square lattices. Their results are good agreement with analytical and experimental results obtained by \cite{foo,K,JA,L,JS,Ga,FT,WGa}. In this work we show that the Zaklan model is very robust for analysis and control of tax evasion, because we use a nonequilibrium dynamics (MVM) to simulate the Zaklan model, that  is the opposite of the study done by \cite{zaklan,zaklan1} equilibrium dynamics (Ising model), and also on various topologies 
\cite{limanew}. Our results are similar to the results obtained by Zaklan et al. \cite{zaklan,zaklan1} regardless of dynamic or topology. As we do not live in a social equilibrium and any rumor or gossip can lead to a government or market chaos, we believe that a non-equilibrium model (MVM) explains better
events of non-equilibrium. Therefore, as the Zaklan model is a sociophysics and econophysics model, we also believe that the best topology used for simulations of this model are until now  the undirected Barab\'asi-Albert and SHP Social networks.

\bigskip
\bigskip
{\bf Acknowledgments}

The author thanks D. Stauffer for many suggestion and fruitful discussions during the
development this work and also for the reading this paper. We also acknowledge the
Brazilian agency CNPQ for  its financial support. This
work also was supported the system SGI Altix 1350 the computational park
CENAPAD.UNICAMP-USP, SP-BRAZIL.

\end{document}